\documentclass[10pt,a4paper]{article}
\usepackage{times}
\begin{document}
\large
\date{ }

\begin{center}
{\Large Reply to 'Comment for "Limits on a nucleon-nucleon
monopole-dipole (axionlike) P,T-noninvariant interaction from spin
relaxation of polarized $^{3}$He" [arXiv:0912.4963], by
A.P. Serebrov'}

\vskip 0.7cm

Yu. N. Pokotilovski\footnote{e-mail: pokot@nf.jinr.ru; tel:
7-49621-62790; fax: 7-49621-65429}

\vskip 0.7cm
            Joint Institute for Nuclear Research\\
              141980 Dubna, Moscow region, Russia\\
\vskip 0.7cm

{\bf Abstract\\}

\begin{minipage}{130mm}

\vskip 0.7cm
It is shown, that the criticism, presented in [1] is based on an elementary
error in calculation of the collision frequency of an atom in a gas with walls
of a container and misunderstanding of the method used in [2] for obtaining
constraints on new short-range spin-dependent forces.

\end{minipage}
\end{center}
\vskip 0.3cm


\vskip 0.2cm


\vskip 0.6cm

 In a recent arXiv preprint \cite{Ser} A.P.Serebrov published critics of the preprint
\cite{me} in which a new limit is presented on the axion-like monopole-dipole
P,T-non-invariant coupling in a range ($10^{-4} - 1$) cm.
 The limit in \cite{me} was obtained from the existing data on the relaxation
rate of spin-polarized $^{3}$He.

 Irrespective of the question if the constraints obtained in \cite{me} are
quantitatively correct in all the interaction range, and if the
method of the energy shift \cite{Zimm} in the UCN magnetic
resonance is more sensitive than the $^{3}$He spin relaxation
method, I would like to show, that the criticism, presented in
\cite{Ser} is based on elementary error in the calculation of the
collision frequency of an atom in a gas with walls of a container
and misunderstanding of the method used in \cite{me} for obtaining
constraints on new short-range spin-dependent forces.

 1. The mean diffusion time of an atom from one wall of a cell to another one,
which is erroneously taken in \cite{Ser} as the time between atom collisions 
with walls is not the time between atom collisions with walls 
and has not direct relation to the problem, at least in this context.

 The wall collision frequency of an atom in a gas confined in a cell of
volume $V$ and surface $S$ is $f_{wall}=vS/4V$, where $v$ is the
atom velocity, independent of the gas density.
 For $v_{^{3}He}\approx 1.5\times 10^{5}$ cm/s and cylindrical cell diam. 5 cm
and length 5 cm $f_{wall}\approx 5\times 10^{4}$ s$^{-1}$,
compared to typical $\sim 50$ s$^{-1}$ in the UCN chamber.

 From this point of view if the spin relaxation in $^{3}$He is caused by the atom
collisions with walls, the depolarization probability "per one collision with
walls" caused by the hypothetic interaction in the polarized $^{3}$He gas is
$w=1/(f_{wall}T_{1})\sim 10^{-12}$.
 Here the longitudinal spin relaxation time $T_{1}$ was taken from
\cite{Rich,Parn} after subtraction of the contribution to the spin
relaxation rate from the bulk dipole-dipole relaxation
\cite{Newb}: the remaining relaxation time is $T_{1}^{rem}=4466\pm
245$ hours for Ref. \cite{Rich}, and $T_{1}^{rem}=2810\pm 146$
hours for Ref. \cite{Parn}.

 Again this value should be compared with typical UCN depolarization
probability per one collision with walls $\sim 10^{-5}$.

 Another interesting figure for comparison: the time spent by particle during
spin relaxation in vicinity of the wall, more exactly at a distance from the
wall not exceeding the searched spin-dependent interaction range $\lambda$.
 For small $\lambda$ it is $T_{1}(\lambda S/V)$, and is $\sim 10^{3}$ s in a
${^3}$He cell and $\sim$0.05 s in the UCN EDM chamber.

 It is seen that if the spin relaxation were induced by simple spin rotation
in the hypothetic spin-dependent field, the $^{3}$He relaxation method would be 
much more sensitive than the method based on the UCN storage.
  
 But these considerations provoked by \cite{Ser} do not have relation to the
the constraints on new spin-dependent interactions \cite{me} from spin
relaxation of polarized $^{3}$He gas.

 2. The depolarization mechanisms of particles in inhomogeneous magnetic
(pseudo-magnetic) field in a dense and very rarefied (UCN) gases
are somewhat different.

 At the very low density, when the free path length between particle collisions
with walls is less than between collisions with other particles in a gas,
the depolarization probability of a particle spin (ultracold
neutron, for example) per one collision with the wall is
determined by the expression \cite{myUCN}
%1
\begin{equation}
w=\frac{V_{0}^{2}<v_{\perp}>^{2}(1-e^{-d/\lambda})^{2}}
{\lambda^{2}\hbar^{2}\Biggl(\frac{<v_{\perp}>^{2}}{\lambda^{2}}
+4\omega_{0}^{2}\Biggr)^{2}},
\end{equation}
where $\omega_{0}=\gamma_{n}B$ is the neutron spin Larmor
frequency in the external field $B$, $\gamma_{n}$ - the
gyromagnetic ratio for the particle, $<v_{\perp}>$ is the averaged
over the particle spectrum normal to the surface particle velocity
component, and the the monople-dipole (axion-like) potential
between the layer of substance and the nucleon separated by the
distance $x$ from the surface is:
%2
\begin{equation}
V(x)=\mp 2\pi g_{s}g_{p}\kappa\lambda N e^{-x/\lambda}(1-e^{-d/\lambda})=
V_{0}e^{-x/\lambda}(1-e^{-d/\lambda})
\end{equation}
where $N$ is the nucleon density in the layer, $d$ is the layer's thickness,
where $g_{s}$ and $g_{p}$ are the dimensionless coupling constants of the
scalar and pseudoscalar vertices (unpolarized and polarized particles),
$\kappa=\hbar^{2}/(8\pi m_{n})$, $m_{n}$ is the nucleon mass at the polarized
vertex,
$\lambda=\hbar/(m_{a}c)$ is the range of the force, $m_{a}$ - the axion mass.

 In a dense gas the rate of spin relaxation in an inhomogeneous magnetic field
of nuclei polarized along z-axis is determined by particle collisions in a gas
and by the gradient of the field \cite{Sche}:
%4
\begin{equation}
\frac{1}{T_{1}^{grad}}=\frac{1}{3}
\frac{(\partial H_{x}/\partial x)^{2}+(\partial H_{y}/\partial y)^{2}}
{H_{z}^{2}}
<u^{2}>\frac{\tau_{c}}{1+(\omega_{0}\tau_{c})^{2}},
\end{equation}
where $<u^{2}>$ is the mean squared velocity of $^{3}$He atoms in a gas,
$\omega_{0}=2\mu H_{z}/\hbar$ is the magnetic resonance frequency in the
magnetic field applied along z-axis, $\tau_{c}$ is the time between
collisions of the $^{3}$He atoms in a gas.
 Relatively rare collisions with walls are ignored in this consideration.

 When spin relaxation is caused by the gradient of spin-dependent potential
$V$, the rate of spin relaxation is
%5
\begin{equation}
\frac{1}{T_{1}^{grad}}= \frac{4}{3}
\frac{(\partial V_{x}/\partial x)^{2}+(\partial V_{y}/\partial y)^{2}}
{(\hbar\omega_{0})^{2}}<u^{2}>\frac{\tau_{c}}{1+(\omega_{0}\tau_{c})^{2}}.
\end{equation}

 I understand that the consideration in \cite{me} was not complete
(the infinite flat cell was considered instead of finite size cell, which is essential
when $\lambda$ is not very small compared to the cell size) and is not
quite correct for small $\lambda$ because of very different reason.

 But critical arguments presented by A.P. Serebrov in \cite{Ser} are not to the
point and contain error.

\end{document}